\documentclass[lettersize,journal]{IEEEtran}
\usepackage{amsmath,amsfonts}
\usepackage{algorithmic}
\usepackage{algorithm}
\usepackage{array}
\usepackage[caption=false,font=normalsize,labelfont=sf,textfont=sf]{subfig}
\usepackage{textcomp}
\usepackage{stfloats}
\usepackage{url}
\usepackage{verbatim}
\usepackage{graphicx}
\usepackage{cite}

\usepackage{lipsum}  
\usepackage{multirow}
\usepackage{booktabs,siunitx}
\usepackage{colortbl}
\usepackage{amssymb}
\usepackage{pifont}
\usepackage{amsfonts}
\usepackage{tabularx}
\usepackage{array, makecell}
\usepackage{xcolor}
\usepackage{graphicx}
\usepackage{kotex}
\usepackage{comment}
\usepackage{soul}
\usepackage{url}

\begin{document}

\title{Musical Word Embedding for Music Tagging and Retrieval}

\author{SeungHeon~Doh, Jongpil~Lee, Dasaem~Jeong, Juhan Nam~\IEEEmembership{Member,~IEEE}
\thanks{The authors are with the Graduate School of Culture Technology, Korea Advanced Institute of Science and Technology, Daejeon 34141, South Korea (e-mail: seungheondoh@kaist.ac.kr; richter@kaist.ac.kr; juhan.nam@kaist.ac.kr).}
}



\maketitle

\begin{abstract}
Word embedding has become an essential means for text-based information retrieval. Typically, word embeddings are learned from large quantities of general and unstructured text data. However, in the domain of music, the word embedding may have difficulty understanding musical contexts or recognizing music-related entities like artists and tracks. To address this issue, we propose a new approach called Musical Word Embedding (MWE), which involves learning from various types of texts, including both everyday and music-related vocabulary.  We integrate MWE into an audio-word joint representation framework for tagging and retrieving music, using words like tag, artist, and track that have different levels of musical specificity. Our experiments show that using a more specific musical word like track results in better retrieval performance, while using a less specific term like tag leads to better tagging performance. To balance this compromise, we suggest multi-prototype training that uses words with different levels of musical specificity jointly. We evaluate both word embedding and audio-word joint embedding on four tasks (tag rank prediction, music tagging, query-by-tag, and query-by-track) across two datasets (Million Song Dataset and MTG-Jamendo). Our findings show that the suggested MWE is more efficient and robust than the conventional word embedding.



\end{abstract}


%
\IEEEpeerreviewmaketitle

\section{Introduction }

The rise of online music streaming services has led to a significant increase in the number of music tracks that are available to users. For instance, Spotify, a popular music streaming service, has a catalog of over 100 million songs\footnote{https://newsroom.spotify.com/company-info/ accessed on Mar 1, 2023.}. 
Users typically access songs by listening to playlists that are recommended based on their listening history or by searching for specific songs using a text query. Music tagging is one of the many computational methods used to recommend or retrieve songs and has been extensively studied in the field of Music Information Retrieval (MIR). This method is popular as it can easily scale up text annotation to encompass diverse musical semantics, and it compensates for problems associated with collaborative filtering, such as popularity bias and cold-start\cite{turnbull2008semantic,prockupEGSCK15ismir,nam2018deep}.




Music tagging is usually approached as a classification task that uses supervised learning to predict multiple tag labels based on the acoustic features of music tracks. Over the last decade, researchers have focused on developing better classification models, primarily based on Convolutional Neural Networks (CNNs), such as fully-connected CNNs\cite{choi2016automatic}, Musicnn\cite{pons2018end}, SampleCNNs\cite{lee2017sample}, Harmonic CNNs \cite{Won2020DataDrivenHF}, and Short-chunk CNNs\cite{won2020evaluation}. These models have shown progressive improvements in performance on large-scale datasets such as the Million Song Dataset (MSD)\cite{bertin2011million}. However, the previous work has a limitation in that the classification models can only predict a fixed set of tag labels that are seen in the training phase. For instance, the models are often trained using the most frequently used 50 tags in benchmark evaluations\cite{won2020evaluation}. This may not be sufficient for real-world scenarios, which may require even more tags that account for the diverse aspects of music.

One way to address this limitation is by representing the tag output using semantically distributed vectors created through a word embedding and associating the dense tag vector with audio embedding through metric learning\cite{choi2019zero,won2020multimodal}. This embedding-mapping approach enables the model to annotate songs with unseen tags or enables users to retrieve songs using an arbitrary text query within the large vocabulary that the word embedding contains. Typically, the word embedding is pre-trained using a large-scale word corpus such as Common Crawl or Wikipedia. Although such general corpora provide a large set of vocabulary, they may lack musical context. For instance, the word ``jungle'' is more likely to be understood as a tropical forest than as a genre of dance music in a general context. To address this issue, Won et al. attempted to train the word embedding using text sources specific to the music domain \cite{won2020multimodal}. They demonstrated that domain-specific word embeddings capture musical context better than general word embeddings. However, this approach did not necessarily improve the retrieval performance, likely because the domain-specific word embedding may be too strongly biased towards musical context and may lack an understanding of the general context in music listening, such as mood or user activity. This suggests that a balanced word embedding that incorporates both general and domain-specific contexts is necessary to encompass diverse semantics.

In this paper, we present a customized word embedding for music tagging called \emph{Musical Word Embedding (MWE)}, by using a broad spectrum of text corpora ranging from general to music-specific words in a systematic manner. We define the \emph{musical specificity} of the corpora as a measure of how specific the semantics of the words is to the songs or how general it is. Using various combinations of text corpora with low to high specificity, we first train the musical word embedding and evaluate it with the tag rank prediction task on both seen and unseen tag datasets. We then incorporate the musical word embedding into an audio-word metric learning framework for music tagging and examine how different setups of supervision between audio and words affect the tagging performance on both seen and unseen datasets. Finally, we demonstrate that the audio-word metric learning model, jointly supervised by tags, artist ID, and track ID through the musical word embedding, outperforms previous work based on a general word embedding.

The remainder of this paper is structured as follows: Section II reviews related work on audio-word joint representation, domain-specific word embedding, and metric learning in the music domain. Section III explains the training of the musical word embedding using both general and music corpora. Section IV describes the audio-word metric learning framework. Section V discusses the experimental results. Finally, Section VI presents our conclusions and outlines future work.


\section{Related Works }

\subsection{Word Embedding}
Classic word embedding methods, such as Word2Vec\cite{mikolov2013distributed} and GloVe \cite{pennington2014glove}, use large-scale text corpora (e.g., Google news, Common Crawl) to capture general semantics in a vector space for natural language processing or other downstream tasks. However, since the meaning of words often changes in different domains, and some domains use highly technical terms or jargon, customized word embeddings have been developed for specific domains. For instance, Zhang et al. introduced BioWord2Vec, a biomedical word embedding that combines subword information from unlabeled biomedical text with a widely-used biomedical controlled vocabulary called Medical Subject Headings (MeSH) \cite{Zhang2019}. In the music domain, Won et al. trained a word embedding using music-domain text data, including Amazon reviews, music biographies, and Wikipedia pages about music theory and genres \cite{won2020multimodal}. They observed that the domain-specific word embedding facilitates capturing musical contexts, particularly sub-genre names with bigrams (e.g., deep\_house', western\_swing'). In this work, we train word embeddings using different combinations of general and music-domain corpora to investigate their effect on music tagging and retrieval tasks.

\subsection{Audio-Word Joint Embedding}
There are various approaches to learning a joint embedding space between audio and words for music tagging and retrieval. One approach is to learn a latent space of tags within the training set and then associate the latent space with the audio embedding using metric learning. For instance, Schindler and Knees used Latent Semantic Indexing (LSI) to project the tags onto a vector space and mapped the vectorized tag to the audio embedding using a triplet network \cite{schindler2019multi}. Another approach is to learn a single word embedding in which audio and words are directly mapped. For instance, Watanabe and Goto represented a song using words from lyrics and ``audio words'' from K-means clustering of Mel-Frequency Cepstral Coefficients (MFCCs) of the audio track \cite{watanabe2019query}. They also added the artist ID to the word corpus, considering the difficulty of conceiving appropriate words as a query from the user side. By considering words, audio words, and artist ID within a song as being in the same context, they learned a multi-modal word embedding and called the music retrieval approach \emph{Query-by-Blending}. The last approach is to use a word embedding trained with a large vocabulary to learn an audio-word joint embedding. For instance, Choi et al. used the GloVe model trained with a large corpus of general words and associated the word embedding of tags with the audio embedding using metric learning \cite{pennington2014glove,choi2019zero}. Our work follows this approach but customizes the word embedding using both a general corpus and a music corpus.

\subsection{Supervision in Metric Learning}
The key to metric learning is to learn an embedding space in which the embedded vectors of similar samples are close to each other, while those of dissimilar samples are far apart. In the field of MIR, an important issue in metric learning is how to supervise the similarity between two music samples. One readily available source of supervision is the metadata of music tracks. Early work by Slaney et al. used album ID, artist ID, and blog IDs to linearly transform acoustic features of songs into a Euclidean metric space \cite{Slaney2008LearningAM}. Later, Park et al. used a triplet loss formed with artist ID to learn a CNN-based embedding space \cite{park2017representation}, and Lee et al. extended the model by jointly training it with artist ID, album ID, and track ID \cite{lee2019representation}. Another source of similarity is from human data, such as surveys or listening history. McFee and Lanckriet trained an embedding model using rank-based artist similarity measured by a web-based survey \cite{McFee2009HeterogeneousEF} or song similarity derived from collaborative filtering based on users' listening data \cite{McfeeEtAl_2010_LearSimiFromColl}. Wolff et al. compared several embedding models using rank-based song similarity obtained from the TagATune game \cite{daniel_wolff_2012_1416600}. Lastly, tag labels can also be used to form similar or dissimilar pairs in metric learning. Lee et al. explored disentangled embedding space using genre, mood, and instrument tags \cite{lee2020disentangled, lee2020metric}. In this work, we use multiple similarity notions, such as artist, track, and tags, for supervision in audio-word metric learning. Additionally, unlike previous work, we use these similarity notions for audio-word joint embedding learning.

\section{Methods}
This section presents the detail of training word embedding and audio-word joint word embedding. 

\begin{figure}[!t]
\centering
\includegraphics[width=\linewidth]{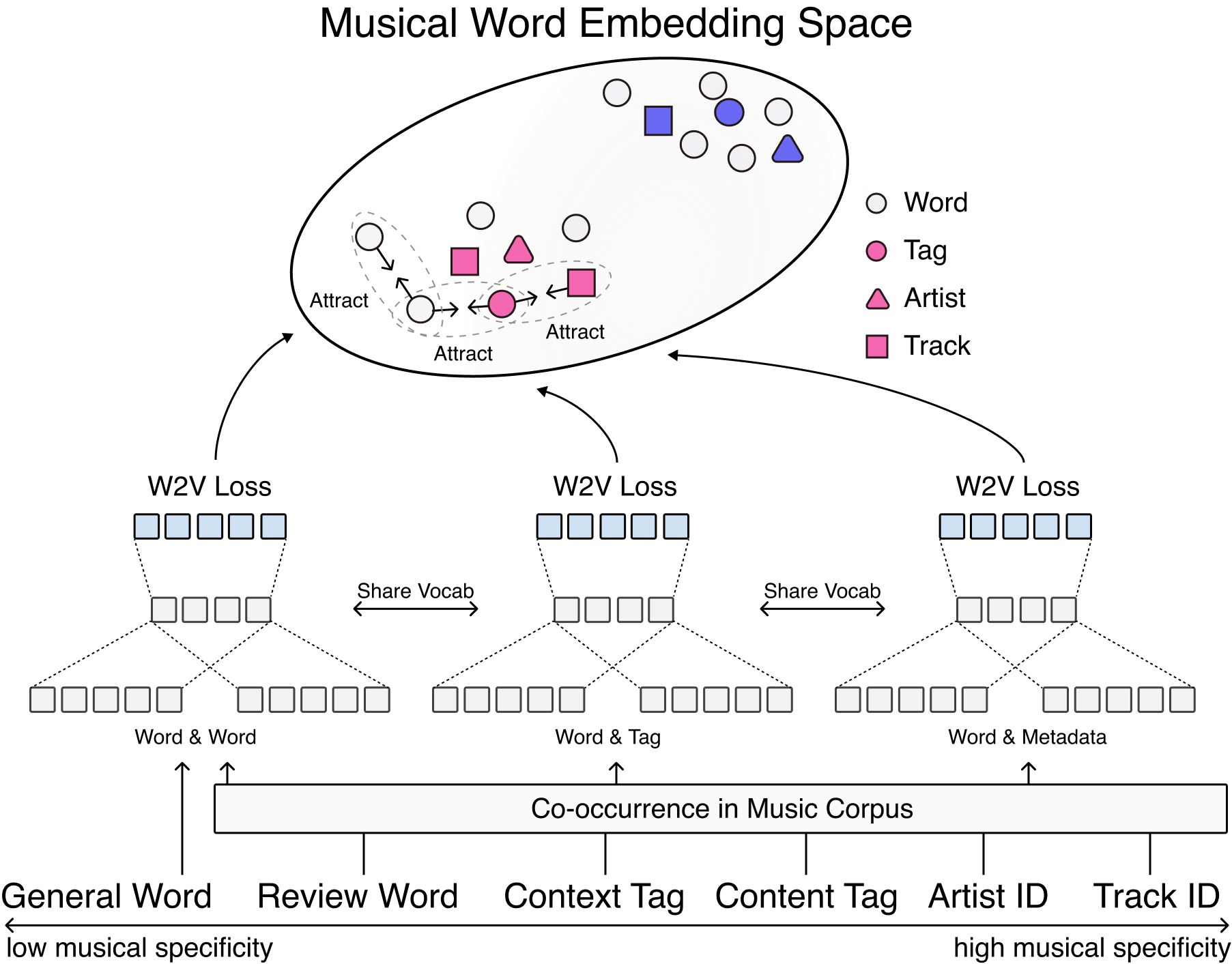}
\vspace{-1mm}
\caption{An illustration of training the musical word embedding. Word embedding vectors within a context window are shown with the same color (pink or blue).}
\vspace{-3mm}
\label{fig_1}
\end{figure}

\subsection{Word Embedding}
\label{sec:MWE}

We trained the word embedding using a wide spectrum of word corpora distributed along the axis of musical specificity. Figure \ref{fig_1}  illustrates the overview of how we trained the musical word embedding. The corpora are mainly divided into a general corpus and a music corpus. The general corpus consists of text documents with a very large vocabulary, such as Wikipedia or Common Crawl. Since the words in the general corpus (i.e., general words) have no specific musical context, they have the lowest musical specificity. The music corpus is a collection of review documents, tags, and artist/track IDs. Review documents describe the backgrounds or musical quality about an artist's album or tracks, covering a large vocabulary while retaining musical semantics. Review words have the second lowest musical specificity on the axis of musical specificity. Tags are categorical labels directly annotated to individual tracks, which we divided into context tags and content tags. Context tags include mood, theme, and usage categories that account for song characteristics from the listeners' perspective. Content tags include genre and instrument categories that are more related to the acoustic characteristics of the songs themselves. Comparing the two types of tags, content tags have higher musical specificity. Artist IDs and track IDs are metadata of music tracks used for music services. Although artist IDs and track IDs have corresponding names, such as "Oasis" (artist name) or "Wonderwall" (track name), we used the index code (or hash code), such as "TRHLXWK128EF35DF13" \footnote{This is an example of an MSD track ID.}, to avoid confusion with general words that have the same spellings. In the MWE, the IDs are regarded as part of the vocabulary. The notion of the artist is generally more specific than that of genre, and therefore, we positioned artist IDs next to content tags. Lastly, track IDs have the highest musical specificity.

Table \ref{tab:1} presents the statistics of the word corpora used to train the MWE in our experiment. The general corpus we used is Wikipedia 2020, which covers approximately 9.8M unique words in the vocabulary. The music corpus is obtained from three publicly available datasets: MuMu, Last.fm, and AllMusic. The MuMu dataset includes music review documents, covering about 660K unique words and 45K artist/track IDs. The Last.fm and AllMusic datasets contain tags and artist/track IDs, with about 1,100 and 1,400 tags, and 32K and 25K artist IDs, and 428K and 507K track IDs, respectively. About 31\% of review words and about 73\% of tags are included in the vocabulary of the general corpus. This overlap enables the bridging of word semantics across the different levels of musical specificity.

To train the word embedding, we calculate the affinity among general words, review words, tags, and IDs within a context window. In the general corpus, the context window is taken directly over sentences in the documents, which is a standard technique in word embedding. For the music corpus, we create a paragraph that is tied to a particular music track by combining the corresponding review document with the tags and artist/track IDs. To balance the data size and blend the review words and artist/track IDs uniformly, we randomly shuffle the paragraph. Specifically, since the total tokens of the general corpus are four times greater than those of the music corpus, as shown in Table \ref{tab:1}, we repeat each sentence in the review document four times and randomly shuffle the word order before combining them with the tags and artist/track IDs to form a paragraph. Next, we take a context window over the shuffled paragraph and learn the affinity among review words, tags, and IDs.

The task of learning dense representations of words is typically achieved through the use of models such as Word2Vec \cite{mikolov2013distributed} or GloVe \cite{pennington2014glove}. Word2Vec has two implementations: continuous bag-of-words (CBOW) and skip-gram. CBOW combines the embeddings of context words to predict the target word, while skip-gram uses the embedding of each target word to predict its context words. In contrast, GloVe trains word embeddings on the non-zero elements in a global word co-occurrence matrix, which can improve the representation of less frequent words. In our work, we used skip-gram to train MWE since it is better at representing less frequent words \cite{mikolov2013efficient}, which is beneficial for capturing musical word semantics. Specifically, given a sequence of training tokens $w_{1},w_{2},...,w_{T}$, the objective of skip-gram is to maximize the average log probability, where $c$ is the size of the context window.



\begin{table}[!t]
\centering
\caption{Statistics of word corpora used to train the musical word embedding}
\vspace{-1mm}
\label{tab:1}
\begin{tabular}{l|c|c|c|c}
\toprule
 & \textbf{General} & \multicolumn{3}{c}{\textbf{Music}} \\  
 & \textbf{Corpus} & \multicolumn{3}{c}{\textbf{Corpus}} \\ 
 \midrule
Data Source & Wikipedia 2020 & MuMu & Last.fm & AllMusic \\ \midrule
Entity Type & Document  & Review, ID & Tag, ID  & Tag, ID \\ \midrule
Entity Number & 4,848,680 & 447,406 & 428,408 & 507,435 \\ 
Unique Track & - & 31,471 & 428,408 & 507,435 \\ 
Unique Artist & - & 14,013 & 32,752 & 25,203 \\ 
Unique Tag & - & - & 1,147 & 1,402 \\ 
Unique Word & 9,868,901 & 660,014 & - & - \\ 
Vocabulary & 9,868,901 & 705,498 & 462,307 & 534,040 \\ \midrule
Total Tokens & 2,746,156,881 & \multicolumn{3}{c}{78,263,644} \\ 
\bottomrule
\end{tabular}
\vspace{-3mm}
\end{table}

\begin{equation}
\begin{aligned}
\label{eqn:eqlabel}
\begin{split}
\frac{1}{T} \sum^{T}_{t=1} \sum^{}_{-c\leq j \leq c, j\neq 0} \log {p(w_{t+j}|w_{t})}
\end{split}
\end{aligned}
\end{equation}

As mentioned earlier, a significant proportion of review words and tags also appear in the Wikipedia corpus and co-occur with artist/track IDs during the training phase. These common words serve as a bridge between general and music-specific semantics.



\begin{figure*}[!t]
\centering
\includegraphics[width=\textwidth]{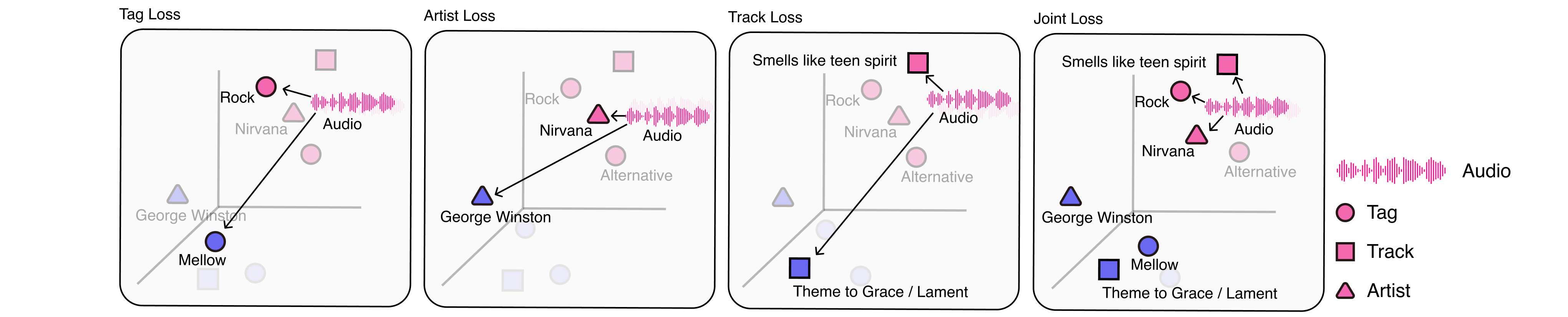}
\vspace{-1mm}
\caption{An illustration of the losses for supervision in the audio-word metric learning. Embedding vectors associated with the anchor are colored in pink if they are of the same class of the anchor (i.e., positive) and in blue otherwise (i.e., negative).} 
\label{fig_loss}
\vspace{-3mm}
\end{figure*}

\subsection{Audio-Word Joint Embedding}

After training the word embedding, we establish a connection to audio embedding by learning a joint embedding space between the two modalities. We adopt a metric learning framework, similar to previous work \cite{choi2019zero}, which learns a similarity score between audio and semantic prototype vectors. Each prototype vector is categorized by tag, artist, and track. We use a triplet network format, where we use two different encoders for each modality. The audio encoder $f(x)$ learns to map the audio mel-spectrogram to the joint embedding space, while the semantic encoder $g(x)$ learns to map semantic information to the joint embedding space. The input to the framework is a triplet of music content items: a query track (anchor), a similar prototype (positive), and a dissimilar prototype (negative) to the anchor. To optimize the network, we use a max-margin hinge loss function, as shown below:
\vspace{1mm}
\begin{equation}
\small
\label{eq1}
\begin{aligned}
\mathcal{L}(A,P)=\text{max}[0, \Delta - Sim(A, P^{+}) + Sim(A, P^{-})]
\end{aligned}
\end{equation}
\vspace{1mm}
where $\Delta$ is the margin, $P^{+}$ denotes the positive prototype vector for the audio input, and $P^{-}$ denotes the negative prototype vector, which is randomly sampled from a set of prototypes without the positive prototype. The cosine similarity of the audio and semantic encoders is used as a similarity function. 




\vspace{1mm}
\begin{equation}
\small
\label{eq2}
\begin{aligned}
Sim(A,P) = \frac {f(A)^{T} \cdot g(P)}{||f(A)|| \cdot || g(P)||}
\end{aligned}
\end{equation}
\vspace{1mm}

We conduct audio-word metric learning on labeled audio datasets. In contrast to prior work \cite{choi2019zero}, our approach allows for the use of both tags and artist/track IDs in the semantic branch. Thus, the total loss function can be expressed as a weighted sum of three loss functions, corresponding to tags, artist IDs, and track IDs, as shown below:



\vspace{1mm}
\begin{equation}
\small
\label{eq3}
\begin{aligned}
\mathcal{L} =  \lambda_\text{Tag} \mathcal{L} (A,P_\text{Tag}) + \lambda_\text{Artist} \mathcal{L}(A,P_\text{Artist}) +  \lambda_\text{Track} \mathcal{L}(A,P_\text{Track})
\end{aligned}
\end{equation}
\vspace{1mm}

Figure~\ref{fig_loss} depicts the joint learning process that combines multiple sources of supervision. We evaluate the performance of the model on music tagging tasks using various combinations of supervisory signals. The audio-word metric learning enables zero-shot learning, since the pre-trained MWE can handle a rich vocabulary beyond the tags and artist/track IDs used during training. This means that the model can accurately predict previously unseen tags and retrieve songs based on arbitrary tags. In our experiments, we also evaluate the model's performance in such a zero-shot learning scenario.




\begin{figure*}[!t]
\centering
\includegraphics[width=\linewidth]{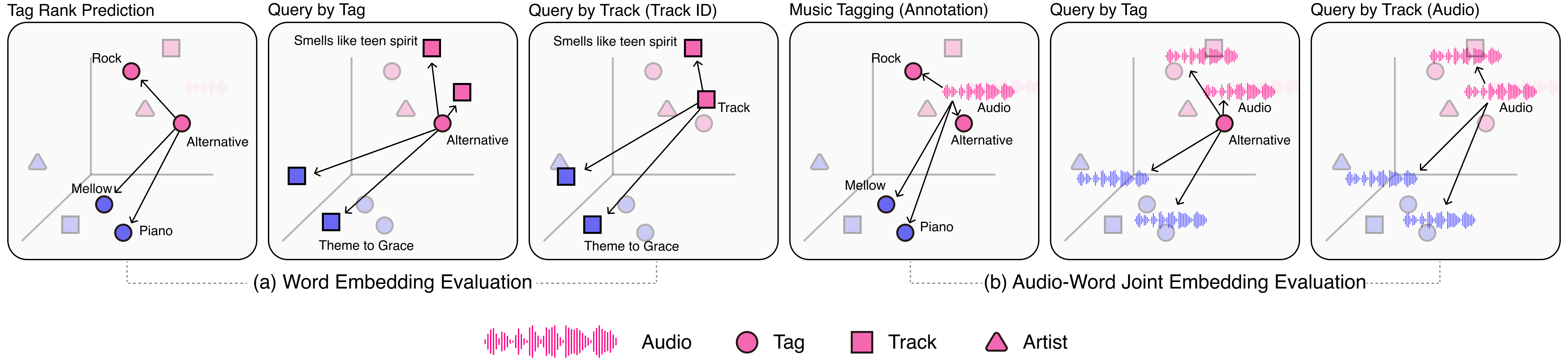}
\vspace{-1mm}
\caption{A summary of evaluation tasks for word embeddings and audio-word joint embeddings. The embedding vectors associated with the input (or query) are colored in pink if they are of the same class (i.e., positive) or in blue otherwise (i.e., negative).}
\label{fig_eval}
\vspace{-3mm}
\end{figure*}

\section{Experiment} \label{Experiment}

\subsection{Datasets}
\subsubsection{Word Embedding}

Table \ref{tab:1} presents the sources of the different types of word corpora used to train our musical word embedding. We obtained the general corpus from Wikipedia 2020\footnote{\url{https://dumps.wikimedia.org/enwiki/20200601/}}, and the music corpus from publicly available datasets: MuMu, Last.fm, and AllMusic. The MuMu dataset\footnote{\url{https://www.upf.edu/web/mtg/mumu}} includes album reviews from Amazon that provide consumer opinions about music\cite{He2016,oramas2017multi}. For tag data, we use annotations from AllMusic and Last.fm. The AllMusic dataset includes content tags (genre, style) and context tags (mood and theme) that were annotated by music experts \cite{schindler2019multi}. The Last.fm dataset contains large crowdsourced tags covering genre, instrumentation, moods, and era. The artist and track ID names were based on those from the Million Song Dataset (MSD) \cite{bertin2011million}. For the music corpus, we clustered the review texts, tags, and artist/track IDs for each audio track using the MSD track ID \footnote{\url{http://millionsongdataset.com/}}. We converted all characters to lowercase and tokenized sentences using whitespace. After pre-processing, our merged corpus contains 9.8M unique words, 37K artist IDs, and 0.7M track IDs.

\subsubsection{Audio-Word Joint Embedding}


We conducted experiments using the audio-word joint embedding in two different scenarios. The first scenario is music tagging, where the same set of tags is used in both the training and test phases. For this scenario, we used 241,889 audio clips from the MSD dataset and the top 50 tags from Last.fm that were annotated on the audio clips, following the common practice in the music tagging task \cite{lee2020metric}. The second scenario is zero-shot learning, where the test phase includes tags that were unseen in the training phase. Following the experiment setting in the previous work \cite{choi2019zero}, we used 406,409 audio clips from MSD and 1,126 tags. We split the tags into 900 seen tags and 226 unseen tags. In our experiment, we used the generalized zero-shot learning test, which evaluates the retrieval performance with all tracks and unseen tags and evaluates the tagging performance with test tracks and all tags \cite{xian2017zero}. Table~\ref{tab:datasplit} summarizes the data split schemes for the two different experiment scenarios.

\begin{table}[!t]
\centering
\caption{Data Split for Audio-Word Metric Learning}
\vspace{-1mm}
\begin{tabular}{ll|c|c}
\toprule
\textbf{Entity} & \textbf{Types} & \textbf{Zero-Shot Learning} \cite{choi2019zero} & \textbf{Music Tagging} \cite{lee2020metric} \\ \midrule
\multirow{3}{*}{Audio} & Train & 349,516 & 201,680 \\ 
 & Valid & 38,836 & 11,774 \\
 & Test & 18,057 & 28,435 \\ \midrule
\multirow{2}{*}{Tag} & Seen & 900 & \multirow{2}{*}{50} \\
 & Unseen & 226 &  \\ \bottomrule
\end{tabular}
\label{tab:datasplit}
\vspace{-3mm}
\end{table}

\subsection{Compared Models and Training Details}

\subsubsection{Word Embedding}
We trained the word embedding using three different configurations: with the general corpus, the music corpus, and both. The last configuration corresponds to the proposed MWE. For all configurations, we used a vector size of 300 and a context window size of 15, and applied the skip-gram method with 15 epochs and 20 negative samples. To preserve track IDs and artist IDs, which typically appear only once or twice, we did not apply frequency-cut-off. Additionally, we compared the three configurations of word embedding with two pretrained word embeddings trained with Common Crawl: one trained using GloVe \cite{pennington2014glove} with 42B tokens, and the other using skip-gram \cite{salle-etal-2016-matrix} with 58B tokens.




\subsubsection{Audio-Word Joint Embedding}
For the audio encoder $f(x)$, we used 3-second audio excerpts represented as a log-scaled mel-spectrogram with 128 mel bins as input. The spectrogram was calculated with a window size of 1024 samples using the Hanning window and a hop size of 512 samples at a sampling rate of 22,050 Hz. The 3-second excerpts were randomly selected from each audio track.


We used the 1D-CNN model developed by Choi et al. \cite{choi2019zero} as the baseline and changed only the word embedding part to evaluate its effect. We used the same hyper-parameter settings, with the audio encoder $f(x)$ consisting of five 1D convolutional layers followed by a batch normalization layer, a rectified linear unit (ReLU) activation, and a max pooling layer. A fixed-size audio embedding vector compatible with the semantic encoder was constructed using an average pooling layer on top of the convolutional layers. The semantic encoder $g(x)$ was constructed using the pre-trained MWE and a fully-connected linear layer. We trained the networks using stochastic gradient descent with a batch size of 128 for 200 epochs, with a 0.9 Nesterov momentum, $1e^{-3}$ learning rate, and $1e^{-6}$ learning rate decay.


In order to benchmark our approach against the current state-of-the-art, we employed the Music Tagging Transformer \cite{won2021transformer}. The Transformed model comprises a CNN front-end and a transformer back-end. The CNN front-end captures local spectro-temporal features, while the transformer globally summarizes the sequence of the extracted features. We removed the last classifier layer and used the model as an embedding extractor, replacing temporal pooling with the special token embedding $\langle$CLS$\rangle$ at the first part of the embedding sequence. We trained the transformer networks using the Adam optimizer with a cosine annealing scheduler, with a batch size of 128 for 200 epochs and $1e^{-3}$ learning rate.

\begin{table*}[!t]
\centering
\caption{Tag rank prediction scores on various word embeddings. \textbf{Ctn} and  \textbf{Ctx} stands for content tags and context tags, respectively. The left side of the arrow is the query tag category and the right side is the target tag category.}
\vspace{-1mm}
\resizebox{\linewidth}{!}{
\begin{tabular}{l|ccccc|ccccc}
\toprule
\multirow{2}{*}{Corpus (Size)} & \multicolumn{5}{c}{AllMusic (nDCG@30)} & \multicolumn{5}{c}{MTG-Jamendo (nDCG@30)} \\
 & Ctn$\rightarrow$Ctn & Ctn$\rightarrow$Ctx & Ctx$\rightarrow$Ctn & Ctx$\rightarrow$Ctx & Average & Ctn$\rightarrow$Ctn & Ctn$\rightarrow$Ctx & Ctx$\rightarrow$Ctn & Ctx$\rightarrow$Ctx & Average \\ \midrule
\multicolumn{11}{l}{{\color[HTML]{9B9B9B} Musical Word Embeddings - SkipGram}} \\
Wiki (Baseline) & 0.135 & 0.133 & 0.071 & 0.259 & 0.150 & 0.416 & 0.474 & 0.321 & 0.566 & 0.444 \\
Wiki + Review & 0.165 & 0.140 & 0.043 & 0.263 & 0.153 & 0.480 & 0.505 & \textbf{0.397} & 0.566 & 0.487 \\ 
Wiki + Tag + IDs & 0.281 & 0.420 & 0.078 & 0.462 & 0.310 & 0.542 & 0.485 & 0.384 & 0.563 & 0.494 \\
Wiki + Review + Tag + IDs & 0.411 & \textbf{0.516} & 0.190 & 0.487 & 0.401 & \textbf{0.553} & 0.504 & 0.390 & 0.541 & \textbf{0.497} \\
w/ Shuffling Augmentation & \textbf{0.529} & 0.460 & \textbf{0.261} & \textbf{0.498} & \textbf{0.437} & 0.492 & \textbf{0.509} & 0.380 & 0.542 & 0.481 \\ \midrule
\multicolumn{11}{l}{{\color[HTML]{9B9B9B} Large-Scale Corpus Word Embeddings}} \\
Common Crawl-GloVe & 0.196 & 0.108 & 0.047 & 0.275 & 0.157 & 0.460 & 0.499 & 0.339 & 0.595 & 0.473 \\
Common Crawl-SkipGram & 0.210 & 0.134 & 0.053 & 0.268 & 0.166 & 0.440 & 0.482 & 0.358 & \textbf{0.599} & 0.470 \\ \bottomrule
\end{tabular}
}
\label{tab:tag_sim}
\vspace{-3mm}
\end{table*}

\subsection{Evaluation Tasks}

Figure~\ref{fig_eval} depicts the music tagging and retrieval tasks we performed in our experiments, using both word embeddings and audio-word joint embeddings. It is worth noting that the latter type of task involves not only tags, track ID, and artist ID, but also audio data.

\subsubsection{Word Embedding}
We evaluate the quality of the word embeddings by measuring the similarity between tags, between tags and tracks, and between tracks themselves, where the tracks are identified by their corresponding track IDs.

\begin{itemize}
    \item Tag rank prediction (tag-to-tag): The quality of word embeddings can be assessed by examining similarity scores for pre-defined relevant word pairs \cite{mikolov2013distributed, Zhang2019}, but in the music domain, there are no established word pairs. To address this, we used the co-occurrence of tags within an audio track as a proxy for manually-annotated word pairs, assuming that tags that share similar semantics tend to appear together on the same track (e.g., electronic, party). We measured the normalized discounted cumulative gain at \textit{30} (nDCG@30) between the sorted tag co-occurrence (ground-truth) and the tag-to-tag similarity of word embeddings (prediction).

    \item Query-by-tag (tag-to-track): We evaluated the ability of the musical word embedding to retrieve track IDs matching a given tag. This task is specific to the musical domain, where track IDs are part of the corpus. We computed the cosine similarity between track IDs and the corresponding tags in the word embedding and treated it as the prediction score. We used the area under the receiver operating characteristic curve for each tag $\text{ROCAUC}_{\text{tag}}$ as an evaluation metric.
    

    \item Query-by-track (track-to-track): We evaluated the performance of our model on the task of retrieving track IDs similar to a given track ID, which can be applied only to musical word embedding where track IDs are part of the corpus. We used recall@K (R@K) based on the tags annotated to the audio tracks \cite{lee2020metric}. Specifically, we first calculated track-to-track similarity using the word embedding and retrieved similar tracks to a given query track. If at least one of the top K retrieved tracks has the same tag labels (e.g., genre, mood, instrument, era) as the query song, the recall@K is set to 1; otherwise, it is set to 0.
\end{itemize}

\subsubsection{Audio-Word Joint Embedding}
Unlike word embeddings, audio-word joint embeddings incorporate audio data and map them to the joint embedding space via the audio encoder. To evaluate the audio-word joint embeddings, we measure the similarity between audio and tag, and between audio and audio.

\begin{itemize}

    \item Music tagging (audio-to-tag): a multi-label classification task that annotates audio tracks with tags. We used clip-wise area under the receiver-operator curve ($\text{ROCAUC}_{\text{clip}}$) as the evaluation metric for this task. To make predictions, we calculated the cosine similarity between the track-level audio embedding and tag embedding, averaged over the audio embedding vectors for a given track.

    \item Query-by-tag (tag-to-audio): the task of retrieving audio tracks that match a given tag. For this task, we measured the cosine similarity between the tag embedding and the track-level audio embedding, and used tag-wise area under the receiver-operator curve ($\text{ROCAUC}_{\text{tag}}$) as the evaluation metric.

    \item Query-by-track (audio-to-audio): the task of retrieving audio tracks similar to a given track. We calculated the similarity between two tracks in the track-level audio embedding space, and used the recall@K (R@K) metric to evaluate performance.  
\end{itemize}

\subsubsection{Zero-Shot Transfer Learning}

To evaluate the zero-shot transfer performance \cite{choi2019transfer}, we used the audio-word joint embedding for a special case of the query-by-tag task. Unlike the zero-shot split query-by-tag task that splits seen and unseen tags from the same dataset (MSD), the zero-shot transfer method uses a different dataset to evaluate the generalization performance. Once the networks were trained, we computed the feature embedding of the audio and tag by their respective encoders. We then calculated the cosine similarity of these embeddings and compared them with the ground truth annotation. The zero-shot transfer evaluation was performed using the MTG-Jamendo \cite{bogdanov2019mtg} datasets, which were fully unseen in the training stage and contained contents and context tags. The MTG-Jamendo dataset includes audio for 55,701 songs and was annotated by 183 different tags covering genres, instruments, and mood/themes. We used the public genre, instrument, and mood/theme splits (split-0) for testing, which included 87 genre tags, 40 instrument tags, and 56 mood/theme tags.




\section{Results: Word Embeddings }
This section presents experimental results with word embeddings in various training settings.

\subsection{Tag Rank Prediction}  
Table \ref{tab:tag_sim} presents the results of tag rank prediction on two different datasets: AllMusic and MTG-Jamendo. The former is a tag dataset used in training the musical word embedding, and thus was used for verifying the training. The latter is an unseen tag dataset used to evaluate the generalization capability of the word embeddings. Both of the tag corpora include several categories, which we divided into content tags and context tags. We then broke down the tag rank prediction into four categories: within-content (Ctn$\rightarrow$Ctn), within-context (Ctx$\rightarrow$Ctx), content-to-context (Ctn$\rightarrow$Ctx), and context-to-content (Ctx$\rightarrow$Ctn). Ctn$\rightarrow$Ctn measures tag similarity under high musical specificity, and we expect this tag rank prediction to be higher for the musical word embedding. Ctx$\rightarrow$Ctx measures tag similarity in a more general sense, and thus we expect this tag rank prediction to not differ much between musical and general word embeddings. Ctx$\rightarrow$Ctn and Ctn$\rightarrow$Ctx reflect how well the context captures musical content and vice versa, and we expect these tag rank predictions to also be higher for the musical word embedding. In AllMusic, we regarded genre and style categories as content tags, and mood and theme categories as context tags. In MTG-Jamendo, we regarded genre and instrument categories as content tags, and mood/theme categories as context tags.

The upper part of Table \ref{tab:tag_sim} compares the tag rank prediction scores on customized word embeddings trained with different combinations of general and music corpus. The word embedding trained with the general corpus (Wiki) serves as the baseline, and we incrementally add reviews, tags, and IDs to the training set. We then apply shuffling augmentation to the entire training set. Adding the music corpus with higher musical specificity (reviews, tags, and IDs) consistently increases the performance in Ctn$\rightarrow$Ctn on both AllMusic and MTG-Jamendo, and the best results are achieved when the entire music corpus is used. However, the shuffling augmentation is not effective on the unseen tag dataset. In Ctx$\rightarrow$Ctx, adding the music corpus with higher musical specificity does not necessarily increase the performance. The tag rank prediction score significantly increases on AllMusic, as expected, but consistently decreases on MTG-Jamendo. This suggests that the word embedding trained with the general corpus already captures the tag similarity in the musical context level well, and thus the information with higher musical specificity (especially tags and IDs) is not beneficial. For Ctn$\rightarrow$Ctx, adding the music corpus improves the overall performance, but adding the tags and ID corpus does not improve the performance on MTG-Jamendo. Interestingly, the review corpus is more beneficial than the tags and ID corpus on MTG-Jamendo, suggesting that the review corpus bridges the semantic gap between content and context well. In Ctx$\rightarrow$Ctn, the performance trend on MTG-Jamendo is similar to that in Ctn$\rightarrow$Ctx. The review corpus plays a role in filling the semantic gap, whereas the tags and ID corpus does not help much. However, the performance trend on AllMusic is different from the other three cases. Adding either the review corpus or the tag/ID corpus to the training set does not significantly affect the performance, but adding both of them creates a synergy that boosts the performance. The average tag rank prediction scores summarize the overall effect of the music corpus. On AllMusic, the score increases proportionally to adding a music corpus with higher musical specificity, and the shuffling augmentation also improves the performance. On MTG-Jamendo, we observe the same trend, but the increment is moderate, and the shuffling augmentation is not beneficial on the unseen tag dataset.

\begin{figure}[!t]
\centering
\includegraphics[width=\linewidth]{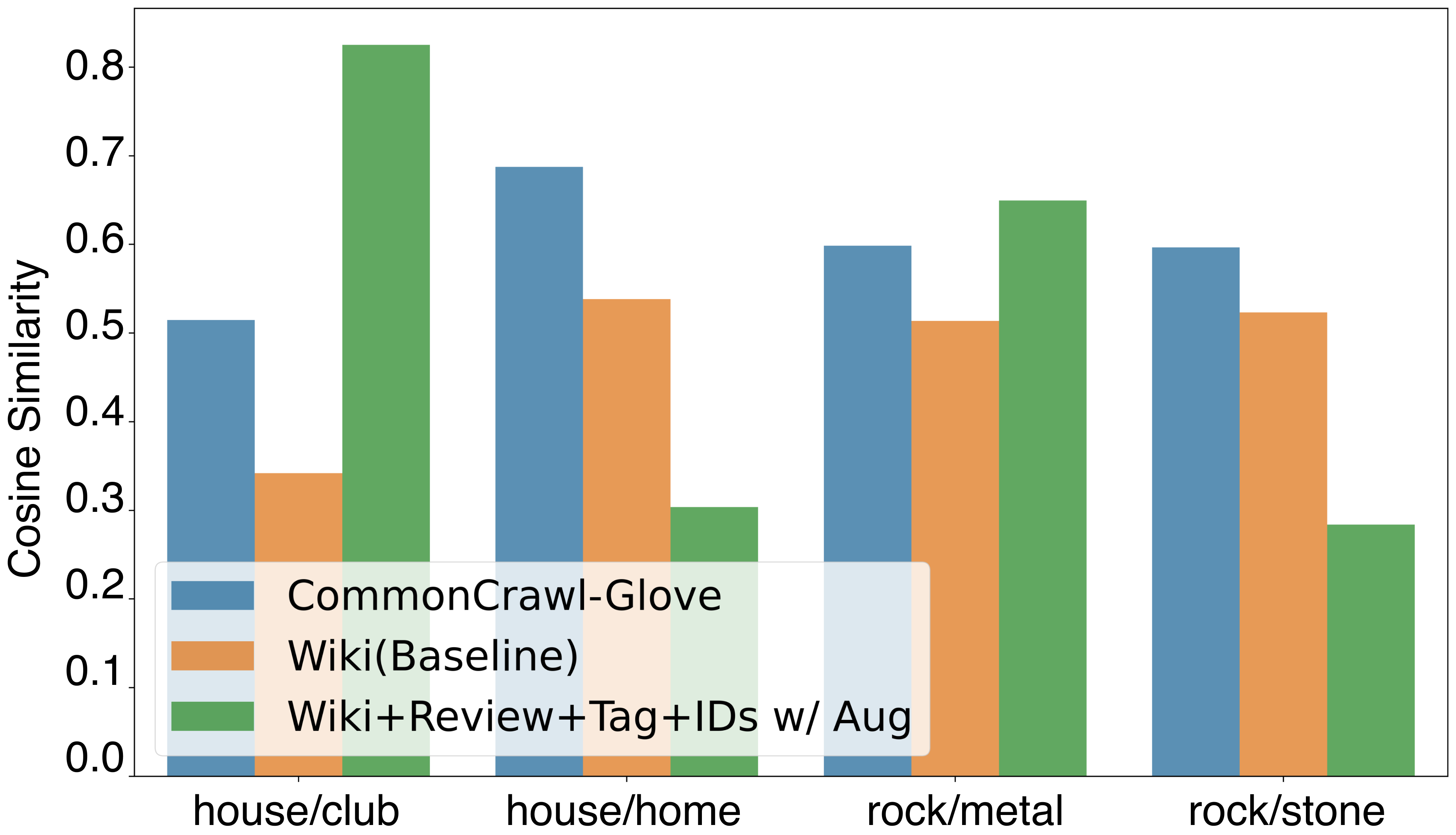}
\vspace{-3mm}
\caption{Comparison of tag cosine similarity between word embedding models.}
\vspace{-3mm}
\label{fig:pairwise}
\end{figure}

The lower part of Table \ref{tab:tag_sim} presents the tag rank prediction scores obtained with pretrained word embeddings. We used two publicly available embeddings trained on Common Crawl, a large-scale general word corpus, with the GloVe and skip-gram methods. The overall performance trends are similar to those obtained with the customized word embeddings trained on the Wiki corpus. The pretrained embeddings capture the similarity of context tags well, particularly on the unseen dataset. However, as the evaluation involves content tags, the tag rank prediction scores are lower than those obtained with the musical word embedding.



Figure \ref{fig:pairwise} presents examples of tag similarity on various word embeddings. It demonstrates that word embeddings trained with a general word corpus, such as Common Crawl and Wiki, have lower cosine similarity in the `house/club' pair than the musical word embedding, since `house' is interpreted as a building rather than a music genre. Conversely, in the `house/home' pair, the musical word embedding has a lower similarity score than the general corpus embeddings. Similarly, the general corpus word embedding exhibits lower cosine similarity in the `rock/metal' pair than the musical word embedding, as `metal' is interpreted as a material rather than a music genre, whereas in the `rock/stone' pair, the opposite result is observed.

\subsection{Query-by-Tag}
Musical word embeddings enable retrieval of music tracks as track IDs are included as part of the corpus used to train the word embeddings. Table \ref{tab:summary} compares the retrieval performance of word embeddings trained with different combinations of general and music corpora. The upper bound of retrieval performance is set by using tags and artist/track IDs in the music corpus, as the word embeddings are concentrated in the region of high musical specificity, resulting in high accuracy in the tag-to-track retrieval task. Adding a general corpus (Wiki) to tags and IDs increases the vocabulary size, but it dilutes musical specificity, which is evident from the reduced performance in tag-to-track retrieval, as shown in Table \ref{tab:summary}. However, adding the review corpus mitigates the performance drop, even with a further increase in the vocabulary size. This suggests that the review corpus effectively bridges the semantic gap between low and high musical specificity. Moreover, the shuffling augmentation technique exhibits significant performance improvement in all cases, as expected, because it increases the sampling frequency of the musical corpus.

\begin{table}[!t]
\centering
\caption{A comparison of query-by-tag performance. `Aug' stands for the shuffling augmentation.}
\begin{tabular}{lcc}
\toprule
 Corpus & Aug & ROCAUC$_{\text{tag}}$  \\  \midrule
\multirow{2}{*}{Tag+ID (Upper bound)} &  & 0.851 \\
 &\checkmark& \textbf{0.930} \\ \midrule
\multirow{2}{*}{Wiki+Tag+ID} &  & 0.673 \\
 &\checkmark& 0.892 \\\midrule
\multirow{2}{*}{Wiki+Review+Tag+ID} &  & 0.800 \\
 &\checkmark& 0.901 \\
\bottomrule
\end{tabular}
\label{tab:summary}
\vspace{-3mm}
\end{table}

\begin{table}[!t]
\centering
\caption{A comparison of query-by-track performance. `Aug' stands for the shuffling augmentation.}
\vspace{-1mm}
\begin{tabular}{lccccc}
\toprule
Corpus & Aug & R@1 & R@2 & R@4 & R@8 \\ \midrule
\multirow{2}{*}{Tag+ID (Upper bound)} &  & \textbf{74.3} & \textbf{82.8} & \textbf{88.8} & 92.8 \\
 &\checkmark& 73.9 & 82.7 & \textbf{88.8} & \textbf{92.9} \\ \midrule
\multirow{2}{*}{Wiki+Tag+ID} &  & 66.0 & 76.7 & 84.7 & 90.3 \\
 &\checkmark& 73.3 & 82.1 & 88.2 & 92.4 \\ \midrule
\multirow{2}{*}{Wiki+Review+Tag+ID} &  & 70.4 & 79.9 & 86.2 & 90.7 \\
 &\checkmark& 73.5 & 82.2 & 88.5 & 92.8 \\
\bottomrule
\end{tabular}
\label{tab:query_by_track_methods}
\vspace{-3mm}
\end{table}

\subsection{Query-by-Track}
Table \ref{tab:query_by_track_methods} compares the performance of track-to-track retrieval using the same sets of word embeddings as in the previous subsection. Once again, we set the upper bound performance using tags and artist/track IDs in the music corpus. We observe that the addition of a general corpus dilutes the retrieval performance, while the subsequent addition of the review corpus alleviates the performance drop. The shuffling augmentation consistently increases the performance for all cases. Interestingly, the recall scores with the shuffling augmentation are similar among the different combinations of corpora. This is likely because the shuffling augmentation ensures that the tags and track IDs are adequately sampled, even when the Wiki or review corpus are added.

\section{Results: Audio-Word Joint Embedding }

Pre-trained word embeddings serve as supervision for training audio-word joint embeddings and as side information for transferring knowledge between seen and unseen classes. The audio-word joint embedding overcomes the limitations of musical word embeddings, which cannot retrieve newly released music, and the limitations of music tagging models, which can only retrieve music with up to 50 tags. This section presents experimental results using audio-word joint embeddings in various training settings (different word embeddings and supervision). In this section, we denote MWE as the customized word embedding trained with the general corpus (Wiki) and the entire music corpus (review, tags, and IDs) along with the shuffling augmentation.


\subsection{Music Tagging and Query-by-Tag }
We first conducted an evaluation of music tagging and query-by-tag (retrieval) performance on MSD using the top 50 tags. Table \ref{tab:tagging_split} compares various audio-word joint embedding models to a classification-based music tagging model \cite{won2021transformer}. While the classification-based model outperforms all audio-word joint embedding models, it can only predict the 50 supervised tags. On the other hand, audio-word joint embedding models can predict not only the 50 tags but also all the vocabularies used in training the word embedding. In our audio-word joint embedding models, we used GloVe as a word embedding and 1D-CNN as an audio encoder as the baseline \cite{choi2019zero}. Replacing GloVe with our MWE increased both ROCAUC\textsubscript{clip} and ROCAUC\textsubscript{tag}, indicating that the musically customized word embedding is more effective than GloVe. We also observed that replacing the 1D-CNN with the Transformer architecture significantly improved the performance. Finally, we used artist/track IDs as additional supervisions for MWE, which includes the IDs as words. Our results show that they consistently increase the tagging performance for both 1D-CNN and Transformer encoders.

\begin{table}[!t]
\centering
\caption{Music Tagging Results on MSD}
\vspace{-1mm}
\resizebox{\columnwidth}{!}{%
\begin{tabular}{lllcc}
\toprule
Audio Model & Word Model & Supervision & ROCAUC$_{\text{clip}}$ & ROCAUC$_{\text{tag}}$ \\ \midrule
\multirow{3}{*}{1D-CNN} & GloVe & Tag & 0.890 & 0.823 \\
 & MWE & Tag & 0.912 & 0.854 \\
 & MWE (Aug) & Tag & \underline{0.918} & \underline{0.863} \\ \midrule
\multirow{15}{*}{Transformer} & GloVe & Tag & 0.927 & 0.869 \\
 & MWE & Tag & 0.926 & 0.867 \\
 & MWE (Aug) & Tag & 0.932 & 0.874 \\
 & MWE & Artist ID & 0.894 & 0.860 \\
 & MWE (Aug) & Artist ID & 0.888 & 0.858 \\
 & MWE & Track ID & 0.892 & 0.868 \\
 & MWE (Aug) & Track ID & 0.892 & 0.870 \\
 & MWE & Tag, Artist ID & 0.929 & 0.869 \\
 & MWE (Aug) & Tag, Artist ID & 0.933 & 0.875 \\
 & MWE & Tag, Track ID & 0.930 & 0.869 \\
 & MWE (Aug) & Tag, Track ID & 0.932 & 0.873 \\
 & MWE & Artist ID, Track ID & 0.907 & 0.866 \\
 & MWE (Aug) & Artist ID, Track ID & 0.896 & 0.872 \\
 & MWE & Tag, Artist ID, Track ID & 0.932 & 0.872 \\
 & MWE (Aug) & Tag, Artist ID, Track ID & \textbf{0.935} & \textbf{0.879} \\
\midrule
 \multicolumn{5}{l}{{\color[HTML]{9B9B9B} Classification Model}} \\
 \multicolumn{3}{l}{Transformer  \cite{won2021transformer}} & - & \textbf{0.892} \\
 \bottomrule
\end{tabular}
}
\label{tab:tagging_split}
\vspace{-3mm}
\end{table}

\subsection{Query-by-Track}
Table \ref{tab:query_by_track_sup} presents the results of the task of retrieving audio tracks similar to a given query track, comparing various audio-word joint embedding models to audio representation models based on the disentangled classification model \cite{lee2020metric}. For the audio-word joint embedding models, we only used the audio encoder since both the query and retrieved results are audio tracks. The overall performance trend is very similar to that in Table \ref{tab:tagging_split}. MWE consistently outperforms GloVe for both 1D-CNN and Transformer audio encoders.



\begin{table}[!t]
\centering
\caption{Query-by-track Results on MSD}
\vspace{-1mm}
\resizebox{\columnwidth}{!}{%
\begin{tabular}{lllcccc}
\toprule
Audio Model & Word Model & Supervision & R@1 & R@2 & R@4 & R@8 \\  \midrule
\multirow{3}{*}{1D-CNN} & GloVe & Tag & 29.6 & 42.9 & 57.2 & 70.7 \\
 & MWE & Tag & 35.9 & 49.7 & 63.9 & 75.5 \\
 & MWE (Aug) & Tag & \underline{38.6} & \underline{52.4} & \underline{65.8} & \underline{76.6} \\ \midrule
\multirow{15}{*}{Transformer} & GloVe & Tag & 44.1 & 57.9 & 70.2 & 80.0 \\
 & MWE & Tag & 41.8 & 55.7 & 68.7 & 78.9 \\
 & MWE (Aug) & Tag & 44.0 & 57.5 & 70.3 & 79.9 \\
 & MWE & Artist ID & 44.0 & 57.7 & 70.1 & 80.2 \\
 & MWE (Aug) & Artist ID & 43.7 & 57.2 & 70.0 & 80.0 \\
 & MWE & Track ID & 44.5 & 58.0 & 70.4 & 80.2 \\
 & MWE (Aug) & Track ID & 44.0 & 57.7 & 70.2 & 80.0 \\
 & MWE & Tag, Artist ID & 43.9 & 57.8 & 70.2 & 80.2 \\
 & MWE (Aug) & Tag, Artist ID & 45.8 & 59.4 & 71.5 & 80.9 \\
 & MWE & Tag, Track ID & 45.0 & 58.7 & 70.7 & 80.1 \\
 & MWE (Aug) & Tag, Track ID & 45.2 & 58.8 & 70.8 & 80.5 \\
 & MWE & Artist ID, Track ID & 46.8 & \textbf{60.4} & \textbf{72.3} & \textbf{81.4} \\
 & MWE (Aug) & Artist ID, Track ID & 46.6 & 59.8 & 71.9 & 81.2 \\
 & MWE & Tag, Artist ID, Track ID & 45.7 & 59.5 & 71.8 & 80.9 \\
 & MWE (Aug) & Tag, Artist ID, Track ID & \textbf{47.1} & 60.2 & 71.9 & 81.2 \\ \midrule
 \multicolumn{7}{l}{{\color[HTML]{9B9B9B} Audio Representation Learning Model}} \\
 \multicolumn{3}{l}{Disentangle Proxy-based Model  \cite{lee2020metric}} & 45.0 & 58.5 & 71.0 & 80.9 \\
 \bottomrule
\end{tabular}
}
\label{tab:query_by_track_sup}
\vspace{-3mm}
\end{table}

\subsection{Evaluation on Zero-Shot Tags}

\noindent \textbf{Comparison with Different Word Embeddings}
The upper section of Table \ref{tab:tag_supervision} presents the zero-shot tagging and retrieval results for three different audio-word joint embedding spaces. Compared to GloVE, MWE shows better performance in the tagging task for unseen audio and the retrieval task for unseen tags. This indicates that the word embedding models trained with a high degree of musical specificity provide better quality supervision for training audio encoders.

\vspace{2mm}
\noindent \textbf{Comparison with Supervisions}
The lower part of Table \ref{tab:tag_supervision} compares the zero-shot tagging and retrieval performance using tag, artist, track, and multiple supervisions. The level of musical specificity increases in the order of tag, artist, and track. When comparing the results between single supervisions, the models trained with track supervision show higher retrieval performance than those trained with tag supervision (0.790 $\xrightarrow{}$0.847 in ROCAUC$_\text{tag}$). On the other hand, the models with tag supervision show higher tagging performance than those with track supervision (0.884$\xrightarrow{}$0.955 in ROCAUC$_\text{clip}$). This is due to the difference in task and musical specificity. The tagging task distinguishes each tag with the given audio, while the retrieval task distinguishes the audio with the given tag. Therefore, when training an audio encoder, tag supervision that discriminates similar and dissimilar tags is suitable for the tagging task, and supervision with high musical specificity, such as artist and track, is suitable for the retrieval task, by discriminating audio more specifically. The model trained by multiple supervisions shows a balanced performance in tagging and retrieval tasks. Comparing both scores, the joint loss model using all three supervisions outperformed the single supervision models.

\begin{table}[!t]
\centering
\caption{Zero-shot Tagging and Retrieval Performance on MSD.}
\vspace{-1mm}
\resizebox{\columnwidth}{!}{%
\begin{tabular}{lllcc}
\toprule
Audio Model & Word Model & Supervision & ROCAUC$_{\text{clip}}$ & ROCAUC$_{\text{tag}}$ \\ \midrule
\multirow{3}{*}{1D-CNN} & GloVe & Tag & 0.904 & 0.679 \\
 & MWE & Tag & 0.941 & 0.747 \\
 & MWE (Aug) & Tag & \underline{0.943} & \underline{0.768} \\ \midrule
\multirow{15}{*}{Transformer} & GloVe & Tag & 0.906 & 0.688 \\
 & MWE & Tag & 0.954 & 0.780 \\
 & MWE (Aug) & Tag & 0.955 & 0.790 \\
 & MWE & Artist ID & 0.911 & 0.813 \\
 & MWE (Aug) & Artist ID & 0.926 & 0.843 \\
 & MWE & Track ID & 0.889 & 0.814 \\
 & MWE (Aug) & Track ID & 0.884 & 0.847 \\
 & MWE & Tag, Artist ID & 0.958 & 0.819 \\
 & MWE (Aug) & Tag, Artist ID & 0.961 & 0.841 \\
 & MWE & Tag, Track ID & 0.953 & 0.826 \\
 & MWE (Aug) & Tag, Track ID & 0.959 & 0.839 \\
 & MWE & Artist ID, Track ID & 0.875 & 0.825 \\
 & MWE (Aug) & Artist ID, Track ID & 0.896 & 0.859 \\
 & MWE & Tag, Artist ID, Track ID & 0.954 & 0.831 \\
 & MWE (Aug) & Tag, Artist ID, Track ID & \textbf{0.959} & \textbf{0.853} \\
\bottomrule
\end{tabular}
}
\label{tab:tag_supervision}
\vspace{-3mm}
\end{table}


\begin{table}[!t]
\caption{Zero-shot Retrieval Performance on MTG-Jamendo}
\vspace{-1mm}
\resizebox{\columnwidth}{!}{%
\begin{tabular}{lll|cc|c}
\toprule
\multirow{2}{*}{Audio Model} & \multirow{2}{*}{Word Model} & \multirow{2}{*}{Supervision} & \multicolumn{2}{c|}{Content} & Context \\ &  &  & Genre & Inst & Mood/Theme \\ \midrule
\multirow{3}{*}{1D-CNN} & GloVe & Tag & \underline{0.794} & \underline{0.564} & 0.618 \\
 & MWE & Tag & 0.782 & 0.504 & 0.626 \\
 & MWE (Aug) & Tag & 0.789 & 0.515 & \underline{0.636} \\ \midrule
\multirow{15}{*}{Transformer} & GloVe & Tag & 0.816 & 0.569 & 0.622 \\
 & MWE & Tag & 0.828 & 0.520 & 0.644 \\
 & MWE (Aug) & Tag & 0.821 & 0.537 & 0.638 \\
 & MWE & Artist ID & 0.827 & 0.556 & 0.662 \\
 & MWE (Aug) & Artist ID & 0.832 & 0.564 & 0.649 \\
 & MWE & Track ID & 0.830 & \textbf{0.596} & 0.647 \\
 & MWE (Aug) & Track ID & 0.838 & 0.590 & 0.661 \\
 & MWE & Tag, Artist ID & 0.840 & 0.524 & 0.660 \\
 & MWE (Aug) & Tag, Artist ID & 0.845 & 0.547 & 0.666 \\
 & MWE & Tag, Track ID & 0.838 & 0.543 & 0.669 \\
 & MWE (Aug) & Tag, Track ID & 0.847 & 0.555 & 0.672 \\
 & MWE & Artist ID, Track ID & 0.829 & 0.571 & 0.664 \\
 & MWE (Aug) & Artist ID, Track ID & 0.838 & 0.594 & 0.657 \\
 & MWE & Tag, Artist ID, Track ID & 0.839 & 0.549 & 0.669 \\
 & MWE (Aug) & Tag, Artist ID, Track ID & \textbf{0.849} & 0.571 & \textbf{0.670}
 \\ \midrule
 \multicolumn{6}{l}{{\color[HTML]{9B9B9B} Audio-Text Representation Learning Model}} \\
 \multicolumn{3}{l|}{TTMR \cite{toward2023doh}} & 0.818 & \textbf{0.669} & 0.601  \\
 \bottomrule
\end{tabular}
}
\label{tab:unseen_eval}
\vspace{-3mm}
\end{table}

\subsection{Zeroshot Transfer Evaluation} 
To evaluate the real-world query-by-tag scenario, we present the $\text{ROCAUC}_{\text{tag}}$ performance using the MTG-Jamendo dataset in Table \ref{tab:unseen_eval}. When using tag supervision, GloVe outperforms MWE in the 1D CNN audio encoder over the genre and instrumental categories, but  MWE outperforms in the deeper transformer audio encoder overall categories. In terms of supervisions, the use of track information, which provides high musical specificity, resulted in higher generalization performance. Notably, the joint supervision with the musical word embedding showed higher performance than the current zero-shot retrieval model using BERT \cite{toward2023doh} in the genre (0.818$\xrightarrow{}$0.849 in ROCAUC$_\text{tag}$) and mood/theme category (0.610$\xrightarrow{}$0.672 in ROCAUC$_\text{tag}$), which demonstrates the effectiveness of our proposed method in real-world scenarios.



\begin{figure*}[!t]
\centering
\includegraphics[width=\linewidth]{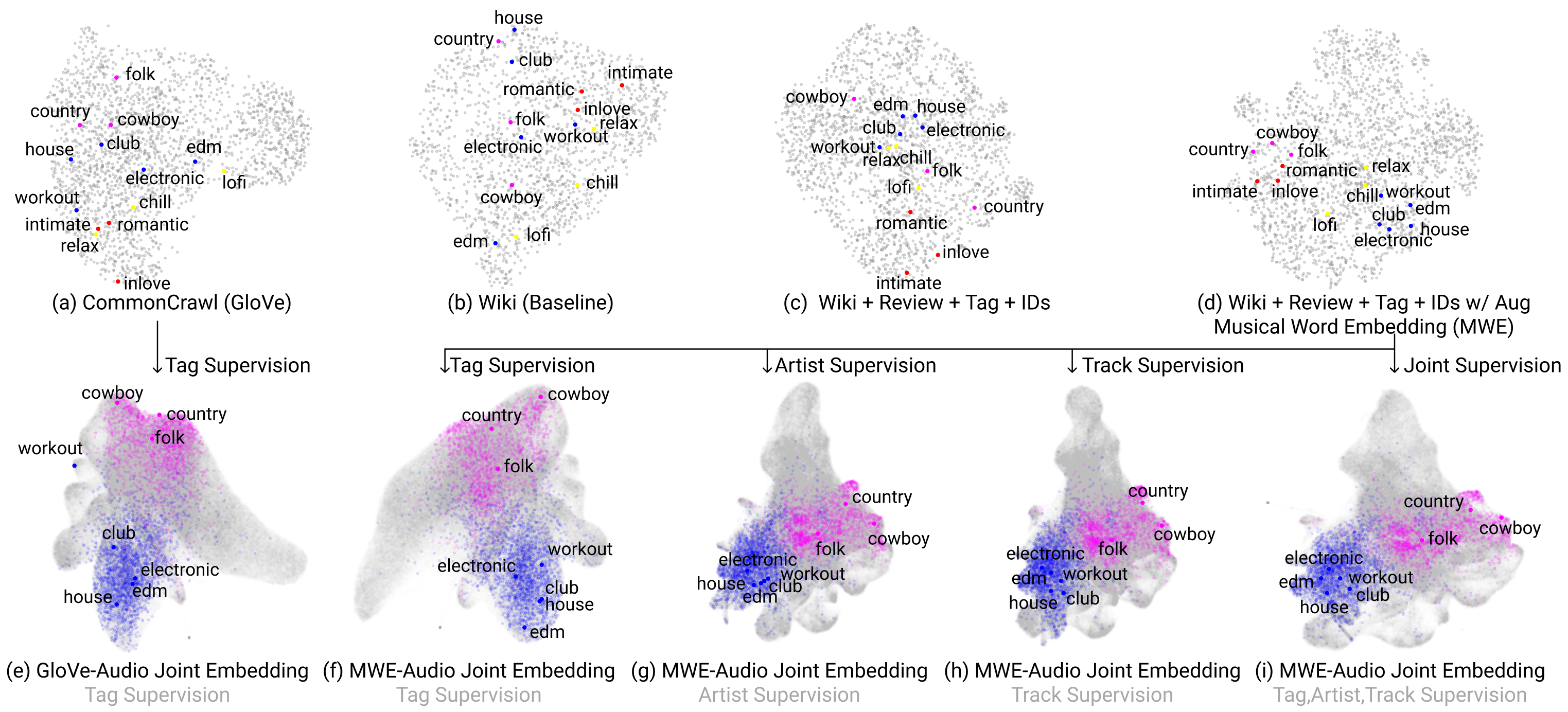}
\vspace{-1mm}
\caption{The UMAP embedding visualization of word embedding (first row) and audio-word joint embedding (second row). Each color represents a similar semantic cluster. We note that (d) is a proposed musical word embedding.}
\vspace{-3mm}
\label{fig:umap}
\end{figure*}

\section{Results: Qualitative Analysis}
This section provides a qualitative analysis of the musical word embedding and audio-word joint embedding by visualization techniques and example-based case studies to broaden the understanding.   

\subsection{Embedding Visualization}
We analyzed the embedding spaces by projecting them into a 2D space using uniform manifold approximation and projection (UMAP) \cite{mcinnes2018umap}. To visualize the word embeddings, we selected 2,201 tag embedding vectors and projected the 300-dimensional vectors into the 2D space. The first row of Figure~\ref{fig:umap} shows the UMAP visualizations of GloVE and MWE. We selected several tags and annotated them with a colored dot and text label. The same color indicates a tag cluster with high similarity in music. For example, `relax', `lofi', and `chill' belong to the same cluster. The two general word embeddings (Fig~\ref{fig:umap}-(a,b)) capture general word similarity well. For instance, emotion tags such as `romantic' and `intimate' are close to each other. However, the majority of tags are more or less scattered, and the same-colored tags are not clustered well. In contrast, the musical word embeddings (Fig~\ref{fig:umap}-(c,d)) show that the same-colored tags are closely located and the clusters are well separated.

To visualize the audio-word joint embedding, we projected the transformer-based joint embedding vectors of all MSD tracks and the 2,201 tag embeddings onto a 2D space. Due to space constraints, we only visualized embeddings trained using GloVe and MWE with augmentation ((d) in the upper row). We selected a few genres and annotated them with colored dots to represent different clusters. The genres related to `electronic', `house', `club', 'edm', and `workout' are colored by blue dots, and those related to `country', `folk', and `cowboy' are colored by magenta dots. 
Comparing the joint embedding space model using tag supervision (Fig~\ref{fig:umap}-(e,f)), the MWE-audio joint embedding space showed stronger cohesion with respect to listening context words such as `club' or `workout' than the GloVe-audio joint embedding space. Additionally, when comparing different supervisions, the artist and track supervision showed stronger cohesion for unseen words such as `cowboy' than the tag supervision (Fig~\ref{fig:umap}-(g,h)). This indicates that the joint embedding space trained with strong musical specificity using artist and track supervision has better generalization than the tag supervision (Fig~\ref{fig:umap}-(i)).

\begin{table}[!t]
\caption{Multi Query Retrieval Results using Musical Word Embedding.}
\vspace{-1mm}
\resizebox{\linewidth}{!}{%
\begin{tabular}{lll}
\toprule
Query (Word) & Top3 Similar Track (MSD\_id) & Track's Annotated Tag \\ \midrule
\multirow{6}{*}{\begin{tabular}[c]{@{}l@{}}deep house \\ in miami ocean\end{tabular}} & Do Ya Like It - Blue 6 & electronica \\ 
 & (TRHXJOS128F426C2D7) & house \\ \cmidrule{2-3} 
 & Trespassers - Newworldaquarium & electronic \\
 & (TRHJNPI128F934C2B4) & club/dance \\ \cmidrule{2-3} 
 & Mama Coca - Jay Haze & \multirow{2}{*}{electronic} \\
 & (TRRGRTL12903CB1F33) &  \\ \midrule
\multirow{6}{*}{\begin{tabular}[c]{@{}l@{}}meditation\\ in the forest\end{tabular}} & Saguaro - Dean Evenson & relax, newage, healing \\
 & (TRVIWIC128F92F9DA8) & ambient, healing.. \\ \cmidrule{2-3} 
 & Ice Castle - Kirsty Hawkshaw & \multirow{2}{*}{ambient} \\
 & (TRHXTEK128F930F2DD) &  \\ \cmidrule{2-3} 
 & InTROsacro - Bruno Sanfilippo & chillout, calm/peaceful \\
 & (TRHLXWK128EF35DF13) & relaxation, ambient... \\
 \bottomrule
\end{tabular}
}
\label{tab:retrieval}
\vspace{-3mm}
\end{table}

\subsection{Music Tagging and Retrieval}
MWE is trained on a combination of Wikipedia, Amazon album review, AllMusic tags, Last.fm tags, and artist/track IDs from MSD. This collection comprises 9.8 million unique general words, 2,201 tags, and 0.7 million tracks for the embedding space. We can retrieve all track items in this space by measuring the similarity score between the text query and the track. If the query contains multiple words, we average the embedding vectors of the words and calculate a similarity score between the query and track. Table~\ref{tab:retrieval} presents multi-query retrieval results using MWE. For instance, when a query such as `deep house in Miami ocean' or `meditation in the forest' is given, MWE interprets `house' as a music genre rather than `home', and understands `forest' or `meditation' as semantically similar to `ambient' or `relax'. Table~\ref{tab:annotation} reports zero-shot tagging results using the joint embedding space. The results are reasonable even if we did not use seen tags in both musical and contextual domains. Further details and demos are available on the website \footnote{\url{https://seungheondoh.github.io/musical_word_embedding_demo/}}.


\begin{table}[!t]
\vspace{-1mm}
\caption{Top 5 auto-tagging results for musical and contextual tag including unseen tags during training.}
\resizebox{\linewidth}{!}{%
\begin{tabular}{cc|cc}
\toprule
\multicolumn{2}{c|}{Nirvana - Smells like teen spirit} & \multicolumn{2}{c}{BTS - Dynamite} \\
Content Tag & Context Tag & Content Tag & Context Tag \\ \midrule
alternativerock & heavy & dancepop & sexy (unseen) \\
hardrock (unseen) & aggressive(unseen) & disco & dance \\
grunge & raucous (unseen) & pop & club (unseen) \\
punkrock & rowdy (unseen) & rnb & clubdance \\
rock (unseen) & angstridden (unseen) & eurodance & party
\\ \bottomrule
\end{tabular}
}
\label{tab:annotation}
\vspace{-3mm}
\end{table}

\section{Conclusions}

This paper introduces the Musical Word Embedding (MWE) model for music tagging and retrieval. MWE leverages a wide range of text corpora, from general to music-specific words, and incorporates the concept of \emph{musical specificity} to measure the level of word semantics related to songs. Our word embedding and joint embedding evaluation demonstrate that the model effectively connects words with varying degrees of musical specificity to songs. Moreover, we have shown potential applications of MWE for music search, including zero-shot music tagging and retrieval. However, our study is currently limited to English language music. Therefore, future work should address multi-lingual music retrieval.

\ifCLASSOPTIONcaptionsoff
  \newpage
\fi



\bibliographystyle{IEEEtran}
\bibliography{IEEEabrv, main.bib}
\end{document}